\documentclass[twoside,twocolumn,9pt]{article}
\usepackage{extsizes}
\usepackage[super,sort&compress,comma]{natbib} 
\usepackage[version=3]{mhchem}
\usepackage[left=1.67cm, right=1.67cm, top=1.785cm, bottom=2.0cm]{geometry}
\usepackage{balance}
\usepackage{times,mathptm}
\usepackage{sectsty}
\usepackage{graphicx} 
\usepackage{lastpage}
\usepackage[format=plain,justification=raggedright,singlelinecheck=false,font={stretch=1.125,small,sf},labelfont=bf,labelsep=space]{caption}
\usepackage{float}
\usepackage{color}
\usepackage{fancyhdr}
\usepackage{fnpos}
\usepackage[english]{babel}
\usepackage{array}
\usepackage{droidsans}
\usepackage{charter}
\usepackage[T1]{fontenc}
\usepackage[usenames,dvipsnames]{xcolor}
\usepackage{setspace}
\usepackage[compact]{titlesec}

\usepackage[colorlinks=true,linkcolor=BluMare,citecolor=Porpora]{hyperref}

\definecolor{cream}{RGB}{90,90,90}
\definecolor{Porpora}{RGB}{157,30,103}
\definecolor{BluMare}{RGB}{25,25,204}

\newcommand{\fsize}{\large}

\newcommand{\blue}{\textcolor{black}}
\usepackage{ulem}

\begin{document}

\pagestyle{fancy}
\thispagestyle{plain}
\fancypagestyle{plain}{

\renewcommand{\headrulewidth}{0pt}
}
%
\makeFNbottom
\makeatletter
\renewcommand\LARGE{\@setfontsize\LARGE{15pt}{17}}
\renewcommand\Large{\@setfontsize\Large{12pt}{14}}
\renewcommand\large{\@setfontsize\large{10pt}{12}}
\renewcommand\footnotesize{\@setfontsize\footnotesize{7pt}{10}}
\makeatother

\renewcommand{\thefootnote}{\fnsymbol{footnote}}
\renewcommand\footnoterule{\vspace*{1pt}%
\color{cream}\hrule width 3.5in height 0.4pt \color{black}\vspace*{5pt}} 
\setcounter{secnumdepth}{5}

\makeatletter 
\renewcommand\@biblabel[1]{#1}            
\renewcommand\@makefntext[1]%
{\noindent\makebox[0pt][r]{\@thefnmark\,}#1}
\makeatother 
\renewcommand{\figurename}{\small{Fig.}~}
\sectionfont{\sffamily\Large}
\subsectionfont{\normalsize}
\subsubsectionfont{\bf}
\setstretch{1.125} 
\setlength{\skip\footins}{0.8cm}
\setlength{\footnotesep}{0.25cm}
\setlength{\jot}{10pt}
\titlespacing*{\section}{0pt}{4pt}{4pt}
\titlespacing*{\subsection}{0pt}{15pt}{1pt}
%
\fancyfoot{}
\fancyfoot[RO]{\footnotesize{\sffamily{1--\pageref{LastPage} ~\textbar  \hspace{2pt}\thepage}}}
\fancyfoot[LE]{\footnotesize{\sffamily{\thepage~\textbar\hspace{3.45cm} 1--\pageref{LastPage}}}}
\fancyhead{}
\renewcommand{\headrulewidth}{0pt} 
\renewcommand{\footrulewidth}{0pt}
\setlength{\arrayrulewidth}{1pt}
\setlength{\columnsep}{6.5mm}
\setlength\bibsep{1pt}
%
\makeatletter 
\newlength{\figrulesep} 
\setlength{\figrulesep}{0.5\textfloatsep} 

\newcommand{\topfigrule}{\vspace*{-1pt}%
\noindent{\color{cream}\rule[-\figrulesep]{\columnwidth}{1.5pt}} }

\newcommand{\botfigrule}{\vspace*{-2pt}%
\noindent{\color{cream}\rule[\figrulesep]{\columnwidth}{1.5pt}} }

\newcommand{\dblfigrule}{\vspace*{-1pt}%
\noindent{\color{cream}\rule[-\figrulesep]{\textwidth}{1.5pt}} }

\makeatother

\twocolumn[
  \begin{@twocolumnfalse}
\vspace{3cm}
\sffamily

\vspace{-1.5cm}
\noindent\Huge{\textbf{Universal contact-line dynamics \blue{at the nanoscale}}} 
\vspace{0.3cm} 

 \noindent\large{Marco Rivetti,\textit{$^{a}$} Thomas Salez,\textit{$^{b}$} Michael Benzaquen,\textit{$^{b}$} Elie Rapha\"el,\textit{$^{b}$} and Oliver B\"aumchen \textit{$^{a}$}} 

 \noindent\normalsize{

The relaxation dynamics of the contact angle between a viscous liquid and a smooth substrate is studied at the nanoscale. Through atomic force microscopy measurements of polystyrene nanostripes we monitor simultaneously the temporal evolution of the liquid-air interface as well as the position of the contact line. The initial configuration exhibits high curvature gradients and a non-equilibrium contact angle that drive liquid flow. Both these conditions are relaxed to achieve the final state, leading to three successive regimes along time: i) stationary-contact-line levelling; ii) receding-contact-line dewetting; iii) collapse of the two fronts. 
For the first regime, we reveal the existence of a self-similar evolution of the liquid interface, which is in excellent agreement with numerical calculations from a lubrication model. For different liquid viscosities and film thicknesses we provide evidence for a transition to dewetting featuring a universal critical contact angle and dimensionless time. 

} 
\vspace{0.3cm}

 \end{@twocolumnfalse} \vspace{0.6cm}

  ]

\renewcommand*\rmdefault{bch}\normalfont\upshape
\rmfamily
\section*{}
\vspace{-1cm}


\footnotetext{\textit{$^{a}$~Max Planck Institute for Dynamics and Self-Organization (MPIDS), Am Fa{\ss}berg 17, 37077 G\"ottingen, Germany. E-mail: oliver.baeumchen@ds.mpg.de}}
\footnotetext{\textit{$^{b}$~Laboratoire de Physico-Chimie Th\'eorique, UMR CNRS 7083 Gulliver, ESPCI ParisTech, PSL Reaserch University, 10 rue Vauquelin, 75005 Paris, France}}




\section{Introduction}

Thin liquid films are ubiquitous in many natural systems and technological applications, spanning from e.g.\ the corneal fluid in the human eye and the aqueous glue in spiders silk to mechanical lubrication, protective coatings, microelectronics fabrication and lithography \citep{eijkel2005nanofluidics, craster2009dynamics, jaeger2001introduction, vollrath1989, shyy2001moving}. 
Understanding the stability and dynamics of thin films is hence a crucial task. 
When a thin liquid layer is deposited on a low-energy surface the substrate might be spontaneously exposed to the vapor phase so to reduce the energy of the system. This phenomenon is achieved either by the nucleation of holes in the film, or by the amplification of capillary waves at the liquid surface \citep{reiter1992dewetting, bischof1996_spinodal, xie1998_spinodal, seemann2001dewetting}, a situation referred to as spinodal dewetting. 
Moreover, in polymer films the confinement of the macromolecules \citep{bodiguel2006reduced, fakhraai2008measuring} or hydrodynamic slip at solid-liquid interface \citep{vilmin2006dewetting, baumchen2009reduced} are also known to influence the mobility and the dynamics of the film. 
%

In some cases, due to the finite size of the film, the interaction of the fluid with the substrate is mediated by the presence of a three-phase contact line where liquid, solid and vapor phases coexist.
This is a common situation for instance in the case of liquid droplets supported by a solid surface. 
It is of primary importance in many applications, e.g.\ ink-jet printing, to know whether the droplets will spread or not on the substrate. In these wetting problems the movement of the three-phase contact line line is  often related to the equilibrium contact angle $\theta_Y$, which follows Young's construction at the contact line: $\cos \theta_Y = ( \gamma_{sv} - \gamma_{sl} ) /\gamma$, where $\gamma_{sv}$, $\gamma_{sl}$ and $\gamma$ are the surface tensions of the solid-vapor, solid-liquid and liquid-vapor interfaces, respectively \citep{degennes2004capillarity, bonn2009wetting}. 
In the vicinity of a moving contact line, however, classical fluid dynamics approaches based on corner flow predict a divergence of the viscous dissipation that would require an infinite force to move the line, as first pointed out by \citet{huh1971hydrodynamic}. 
This apparent paradox can been solved by including microscopic effects like, for instance, slip at the solid-liquid boundary~\citep{dussan1976slip}, the presence of a precursor film ahead the line~\citep{degennes1985wetting} or a height dependence of the interfacial tension~\citep{pahlavan2015thin}. 

A well-established system is that of a liquid droplet spreading onto a complete wettable substrate.  
The so-called Tanner's law predicts the growth of the drop radius with a power law evolution \citep{tanner1979spreading}. 
Surprisingly, this law is valid for every liquid that wets the substrate, and this universality is related to the presence of a thin precursor film ahead of the contact line \citep{degennes1985wetting, bonn2009wetting}. Recently, it has been proven that the power law changes in case of spreading on a thicker liquid layer \citep{cormier2012}. 
Different evolutions are known for droplets on partially wettable substrates. In particular, an exponential relaxation to the equilibrium contact angle is observed when $\theta_Y$ is small and the system is close to equilibrium \citep{deruijter1999, narhe2004contact, ilton2015}.
Tanner's law as well as other investigations of wetting and contact line dynamics are limited to spherical and cylindrical droplets, i.e.\ configurations in which the curvature of the liquid-air interface and therefore the liquid pressure are constant. 
%
More intriguing and complex phenomena may appear in the presence of non-constant film curvature. In the last few years, 
in particular, there has been an increased interest in the capillary-driven dynamics of relaxation of stepped and trench-like film topographies~\citep{mcgraw2011capillary, rognin2011viscosity, mcgraw2012, salez2012, baeumchen2013, mcgraw2013}. In these situations the gradients in Laplace pressure drive a capillary flow mediated by viscosity, leading to the levelling of the interface. 
All these studies are however limited to pure liquid-air interfaces, i.e.\ in the absence of a three-phase contact line. Nevertheless, we note that the latter aspect was recently considered in the context of studies about residual stress~\citep{Guo2014}.

Here, we are studying the relaxation dynamics of the contact angle at the nanoscale by means of tracking the evolution towards equilibrium of a viscous nanofilm in presence of a three-phase contact line. The specificity of this system is the necessity to relax simultaneously both the contact angle and the curvature of the liquid-vapor interface. The resulting liquid dynamics is driven by both Laplace pressure gradients and the balance of forces at the three-phase contact line, while mediated by viscosity. We carry out experiments involving polystyrene films on silicon wafers, a common system that mimics a partial wetting situation \citep{seemann2001polystyrene}. In our experiments the thin films are invariant in one direction and exhibit a rectangular cross section (c.f.\ Fig.\,\ref{fig:stripe}). The spatial and temporal dynamics of the liquid-air interface $z = h(x,t)$ is monitored and distinct regimes corresponding to different mechanisms of relaxation are observed. We show that in general the advancing or receding of the contact line can not be predicted by the simple observation of the initial contact angle, as in a spherical or cylindrical droplet, and propose a geometrical and an energetic approach to describe the evolution of the interface.

\section{Materials and methods}


Polystyrene (PS) thin films are obtained after a spin-coating process and a transfer on Si wafers, following the technique reported in \citet{mcgraw2011capillary}. 
In brief, a solution is prepared by dissolving monodisperse PS (PSS, Germany) in toluene (Sigma-Aldrich, Chromasolv, purity $>99.9\%$). 
Three different molecular weights have been used ($M_\mathrm{w} =$ 3.2, 19 and 34\,kg/mol) 
and solutions were typically prepared with concentration in the range of $1 - 4\%$ in weight. 
The PS solution is spin-coated on freshly cleaved mica sheets (Ted Pella, USA).
During the spin-coating the solvent quickly evaporates resulting in a thin film of the dissolved polymer.
The film is then floated at the surface of ultra-pure (MilliQ) water where
it spontaneously breaks into small pieces. 
Pieces are transferred on  $1 \times 1$ cm$^2$ Si wafers (Si-Mat, Germany), exhibiting a native oxide layer. 
For the purpose of testing the influence of the substrate, Si wafers exhibiting a thick (150\,nm) oxide layer have been used as well. 
Prior to the transfer all Si wafers were cleaned by exposing the substrates to a mixture of hydrogenperoxide and sulfuric acid (``piranha solution''), followed by careful rinse with boiling ultra-pure water \citep{neto2003satellite}.

Once the preparation is completed, a nanostripe exhibiting straight edges is identified with an optical microscope and scanned with an atomic force microscope (AFM, Bruker, Multimode) in tapping-mode.
The sample is then annealed above the glass transition temperature of PS on a high-precision heating stage (Linkam, UK) to induce flow. 
The annealing temperatures are set to 110\,$^\circ$C for the 3.2\,kg/mol molecular weight and to 140\,$^\circ $C or $150\,^\circ$C for the two others. 
After quenching down the PS at room temperature the height profile is scanned with AFM. 
This procedure is repeated several times so to record the temporal evolution of the profiles. 
As an alternative to this \textit{ex-situ} technique, we also performed \textit{in-situ} measurements 
in which the sample is annealed directly on a high-temperature scanner and the liquid interface is monitored by the AFM tip. This way it has been safely ensured that the quenching has no influence on the shape of the profiles.

%

\section{Results}
\subsection{Relaxation of a rectangular interface}

\begin{figure}[t!]
\begin{center}
\setlength{\unitlength}{1cm}
\begin{picture}(8.8,8.4)
\put(1.02,1.){\includegraphics[width=7.8cm]{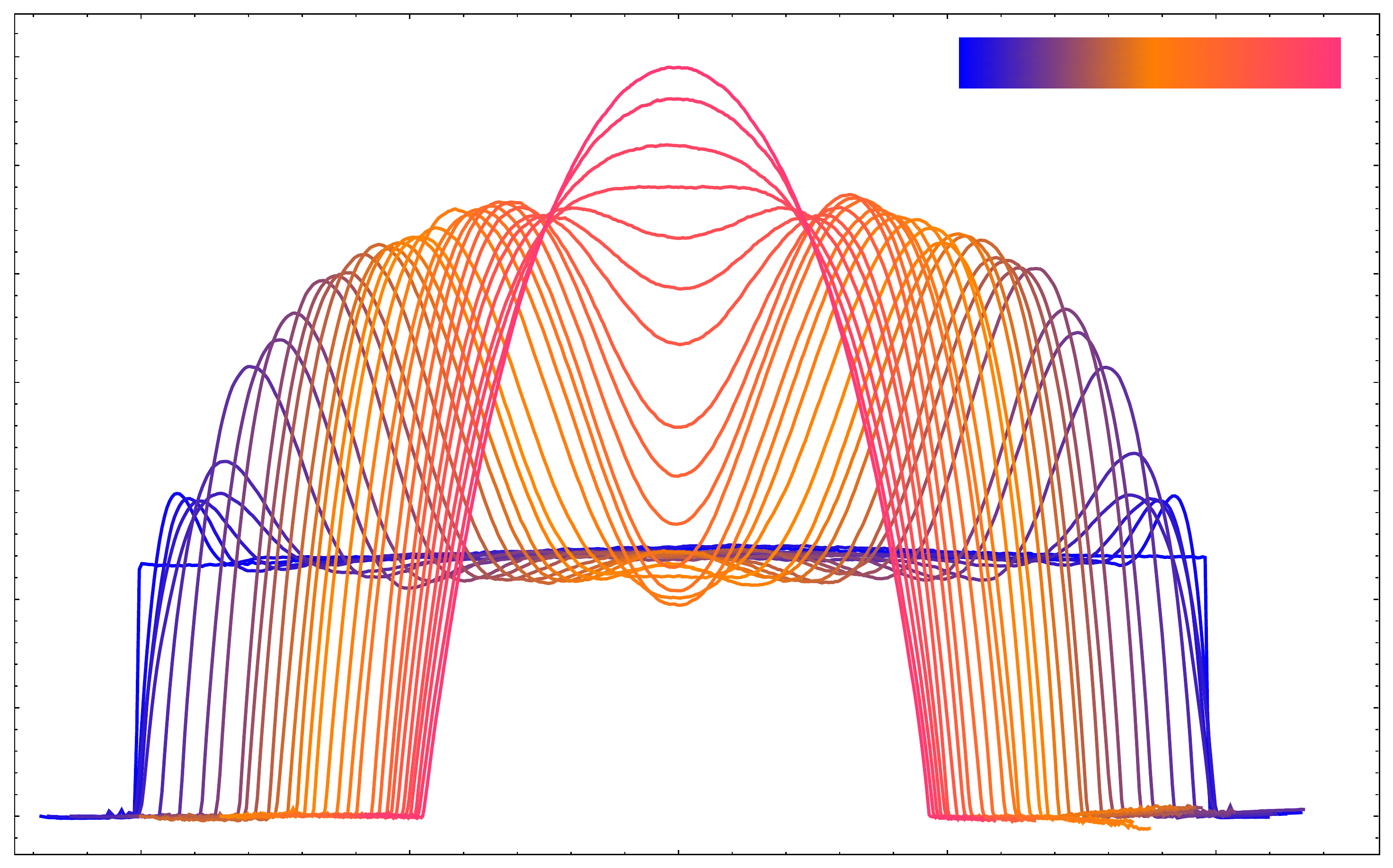}}
\put(0.94,6.3){\includegraphics[width=8.1cm]{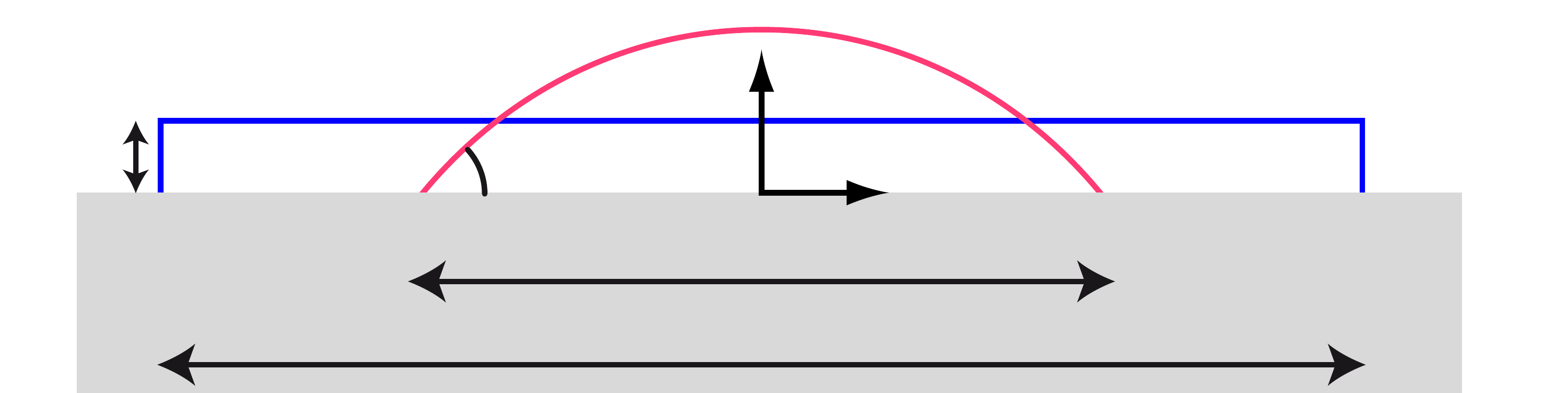}}
\put(4.5,0.2){\Large{$x \; \left[ \mathrm{\mu m} \right]$}}
\put(0.,3){\rotatebox{90}{\Large{$h \; \left[ \mathrm{nm} \right]$} }}
\put(0.8,1.2){\fsize 0}
\put(0.45,2.4){\fsize 100}
\put(0.45,3.6){\fsize 200}
\put(0.45,4.8){\fsize 300}
\put(4.7,0.7){\fsize 0}
\put(6.1,0.7){\fsize 10}
\put(7.6,0.7){\fsize 20}
\put(3.,0.7){\fsize -10}
\put(1.5,0.7){\fsize -20}
\put(7,5.05){\small $t \; \left[ \mathrm{min} \right]$}
\put(6.3,5.05){\small 0}
\put(8.05,5.05){\small 1100}
%
\put(0.15,5.5){\Large (b)}
\put(0.15,8){\Large (a)}
\put(4.65,6.95){\fsize $2 \, r$}
\put(4.6,6.55){\fsize $2 \, \ell_0$}
\put(3.5,7.4){\fsize $ \theta_Y$}
\put(5.5,7.45){\fsize $x$}
\put(5,7.96){\fsize $z$}
\put(1.2,7.45){\fsize $h_0$}
\end{picture}
\end{center}
\caption{(a) Schematic view of the system: the initial rectangle has width $2 \ell_0$ and height $h_0$, while the final cylindrical cap has a contact line radius $r$ and a contact angle $\theta_Y$. 
(b) Series of AFM height profiles displaying the temporal evolution of an initially rectangular PS thin film towards a cylindrical shape ($M_\mathrm{w} = 34$ kg/mol, $T = 140^\circ$C). In the first four profiles the contact line is stationary (within the resolution of the scan), then the dewetting transition takes place and the line recedes. Eventually the coalescence of the two sides leads to a cylindrical cap. Note that the vertical axis is stretched for clarity and the system is invariant in the third direction.  }
\label{fig:stripe}
\end{figure}
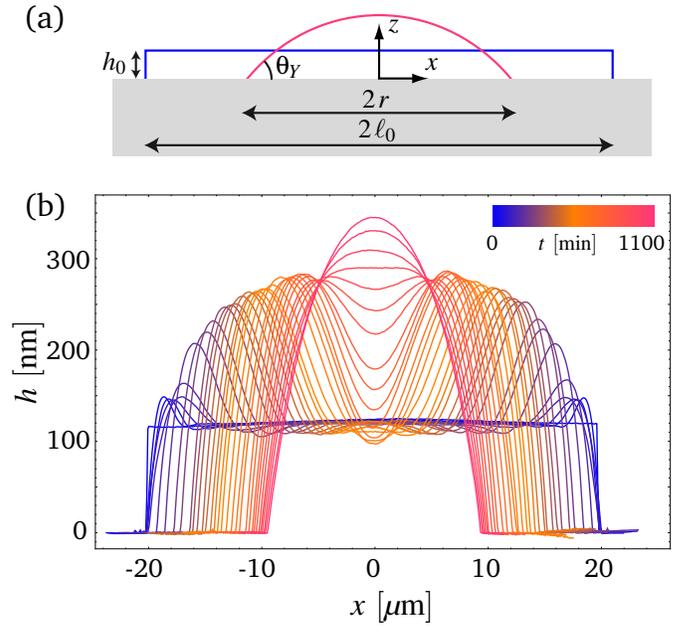


Figure~\ref{fig:stripe} displays a series of profiles corresponding to the temporal evolution of a rectangular interface (the system is invariant in the third dimension) having initial width $2 \ell_0 = 40 \; \mathrm{\mu}$m and a film thickness $h_0 = 120 \; $nm. In the first measure, recorded after 2 min of annealing, we observe that the high Laplace pressure in the corners is rapidly relaxed which gives rise to the formation of a pair of bumps. Note that the vertical axis in the AFM profiles has been stretched in order to have a clear visualization of the details of the interface; thus, the actual slopes of the liquid interface are very small. 
In the following profiles (scanned after 5, 15 and 50 min of annealing, blue lines) the maxima of the bumps slowly move towards the center of the system while both bumps are becoming broader and their height is constant. We note that for these early time profiles no displacement of the contact line is apparent in the AFM data\footnote{Even for the smallest scan sizes of $2 \times 2$\,$\mu m$ and in the presence of a unique reference point, such as a defect on the substrate, no significant movement of the contact line has been detected. Nevertheless, given the limited lateral resolution of the AFM, a displacement of the contact line on the scale of a few nm can not be safely excluded.}. 

For the next profile, at $t= 90$ min, it is clearly visible that both contact lines have receded. We also observe that the bumps have grown due to the accumulation of the liquid that has moved. The retraction of the contact lines continues in the successive scans and the liquid keeps accumulating in the rims that become higher and larger. The velocity of the contact lines in this stage is roughly constant, in agreement with earlier observations by \citet{redon1991dynamics} in the presence of no-slip boundary conditions. 

Around $ t \simeq 700$\,min the front retraction reaches the unperturbed region in the middle of the film and the rims start to merge. The portion of positive curvature disappears and the interface slowly converges to a cylindrical cap. The velocity of the contact lines slows down during the merging process.  Note that at a more advanced stage of the process the system might eventually undergo a Plateau-Rayleigh instability and lose its invariance in the third dimension.

To summarise, from this experiment three different regimes can be identified: i) the initial stationary contact line (SCL) regime is followed by a dewetting transition, as evidenced by ii) the receding contact line (RCL) regime in which the two sides of the nanostripe retract independently, and eventually by  iii) the coalescence regime where the two rims merge to form the cylindrical droplet.  
Note that all along the process the shape of the system remains symmetric. 

At first glance the apparent absence of early spreading, the fixed position of the line at the beginning, and the following dewetting process may sound counterintuitive and surprising. 
However, the perfect parity of the profile in Fig.~\ref{fig:stripe} strongly suggests that these features are general, as opposed to pinning of the contact line on random defects. In order to check the validity of the previous observations, a series of experiments has been carried out involving films with various thickness and viscosities. Due to the parity of the profiles, in the next paragraphs we focus on one single edge of the film and discuss in detail the dynamics of the SCL regime and the dewetting transition. 



\subsection{Stationary Contact Line Regime} 

\begin{figure}[ht]
\begin{center}
\setlength{\unitlength}{1cm}
\begin{picture}(8.8,6.2)
\put(0.5,0.6){\includegraphics[width=8.2cm]{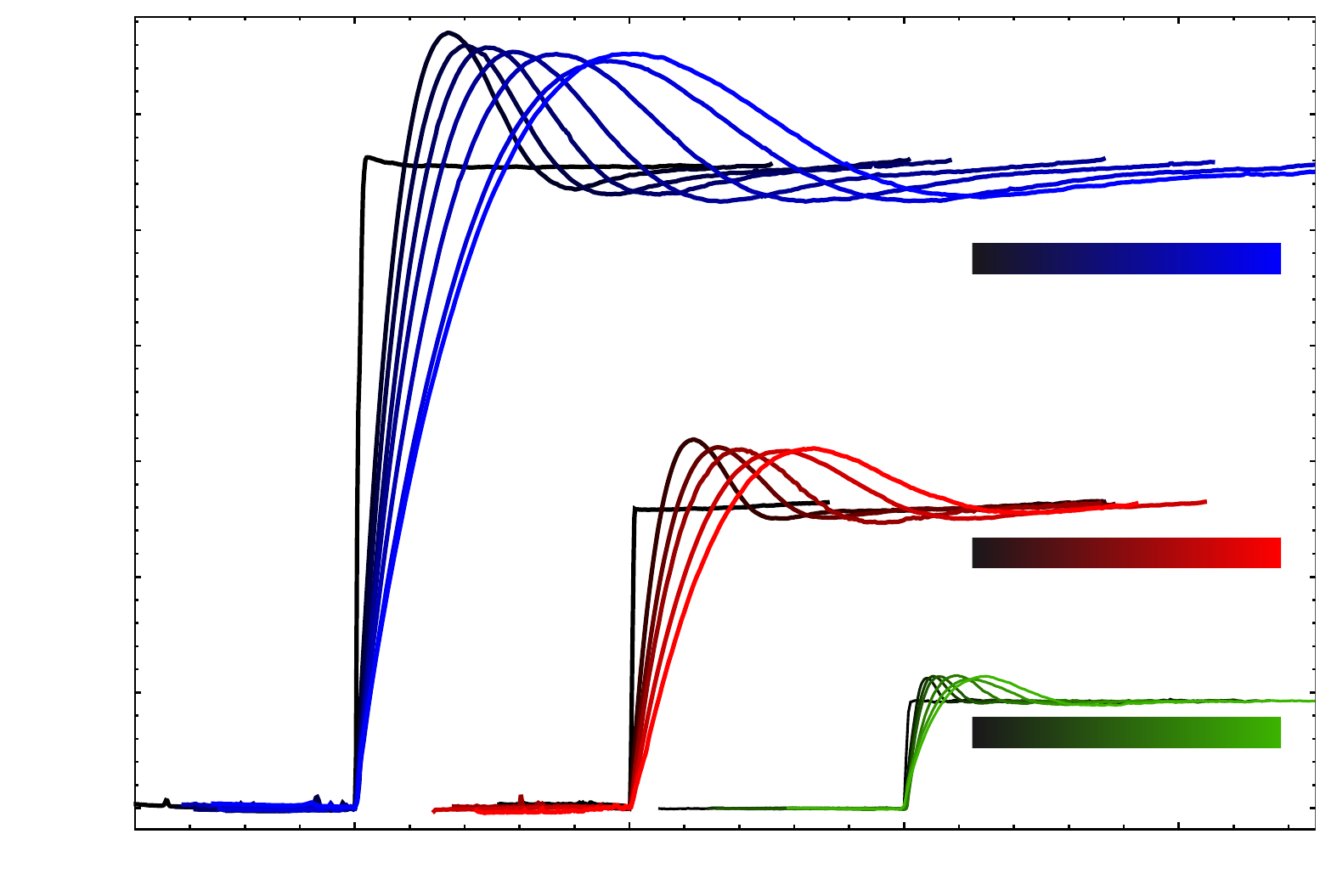}}
%
\put(4.2,0.15){\Large{$x \, \left[ \mathrm{\mu m} \right]$}}
\put(0.15,2.7){\rotatebox{90}{ \Large{$h \, \left[ \mathrm{n m} \right] $ }}}
\put(0.7,5.35){\fsize 300}
\put(0.7,3.9){\fsize 200}
\put(0.7,2.45){\fsize 100}
\put(1.1,1.05){\fsize 0}
\put(2.4,0.65){\fsize -10}
\put(4.32,0.65){\fsize 0}
\put(5.9,0.65){\fsize 10}
\put(7.65,0.65){\fsize 20}
\put(7.1,4.13){\small $t \; \left[ \mathrm{min} \right]$}
\put(6.5,4.13){\small 0}
\put(8.2,4.13){\small 90}
\put(7.1,2.36){\small $t \; \left[ \mathrm{min} \right]$}
\put(6.5,2.36){\small 0}
\put(8.1,2.36){\small 140}
\put(7.1,1.2){\small $t \; \left[ \mathrm{min} \right]$}
\put(6.5,1.2){\small 0}
\put(8.1,1.2){\small 400}
\end{picture}
\end{center}
\caption{Early profiles corresponding to the stationary contact line regime in three experiments involving different film thicknesses and viscosities. 
The experiments are spaced horizontally by 10 $\mu$m for clarity. 
Left (blue) experiment has $h_0 = 280$ nm and $M_\mathrm{w} = 19$ kg/mol at $T = 140\, ^\circ$C, 
middle (red)  has $h_0 = 130$ nm and $M_\mathrm{w} = 3.2$ kg/mol at $T = 110\, ^\circ$C,
right (green)  has $h_0 = 45$ nm and $M_\mathrm{w} = 34$ kg/mol at $T = 140\, ^\circ$C.
}
\label{fig:summary}
\end{figure}
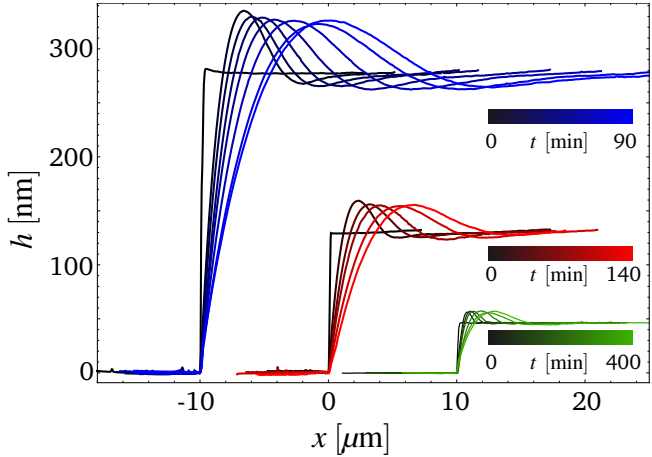

The early evolution of the viscous nanostripe in three different experiments is shown in Fig.~\ref{fig:summary}. All profiles have been recorded before the dewetting transition takes place and corroborate the finding that the contact line is not moving significantly. 
The high Laplace pressure of the initial corner creates a bump in the liquid interface. The width of the bump increases in time while it appears that the height is constant in each experiment. 

The formation of a bump in the presence of a corner as well as the relaxation of the interfacial slope have been studied in the levelling of a stepped interface in a thin viscous film \citep{mcgraw2012, mcgraw2013}. The relaxation of the rectangular interfaces in Fig.~\ref{fig:summary} can be understood in terms of the flow generated by Laplace pressure gradients. As a consequence of the high viscosity of PS and small thickness of the film, the Reynolds number of the flow is very small and inertial effects can be neglected. The liquid dynamics can be safely described using Stokes equation
 $\nabla p = \eta \nabla ^2 \mathbf{v}$,
where $\mathbf{v}$ is the liquid velocity and where the pressure $p$ is related to the curvature of the liquid-air interface by Laplace equation $p = - \gamma \, \partial^2 h / \partial x^2$, only valid for a 2D interface within the small slopes approximation.

Following the theoretical framework summarized in \citet{oron1997long}, the Stokes equation can be further simplified introducing the lubrication approximation. The equation governing the evolution of the liquid interface $h(x,t)$, in the presence of a no-slip boundary condition at the solid-liquid interface, a no-shear boundary condition at the free interface, and in the absence of disjoining forces, can be deduced:
\begin{equation}
\frac{\partial h}{\partial t} + \frac{\gamma}{3 \eta} \frac{\partial}{\partial x} \left( h^3 \frac{\partial^3 h}{\partial x^3} \right)  =  0 \;.
\label{eq:tfe}
\end{equation}
It can be proven~\citep{huppert1982propagation, aradian2001} that this equation admits self-similar solutions of the form $h(x,t) = h (x / t^{1/4})$. This self-similiarity has been verified experimentally for different geometries \citep{mcgraw2012, mcgraw2013, baeumchen2013, chai2014}, all of them being limited to pure liquid-air interfaces. We now check whether this self-similarity holds in the presence of a fixed contact line.

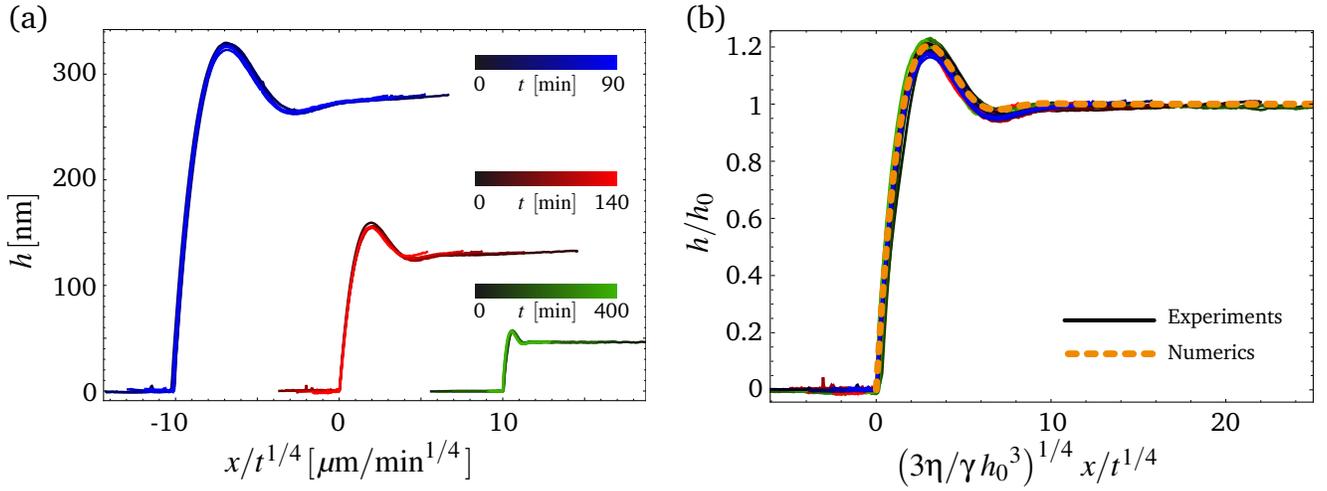
\begin{figure*}
\begin{center}
\setlength{\unitlength}{1cm}
\begin{picture}(17.5,6.45)
%
\put(0.6,0.7){\includegraphics[height=5.5cm]{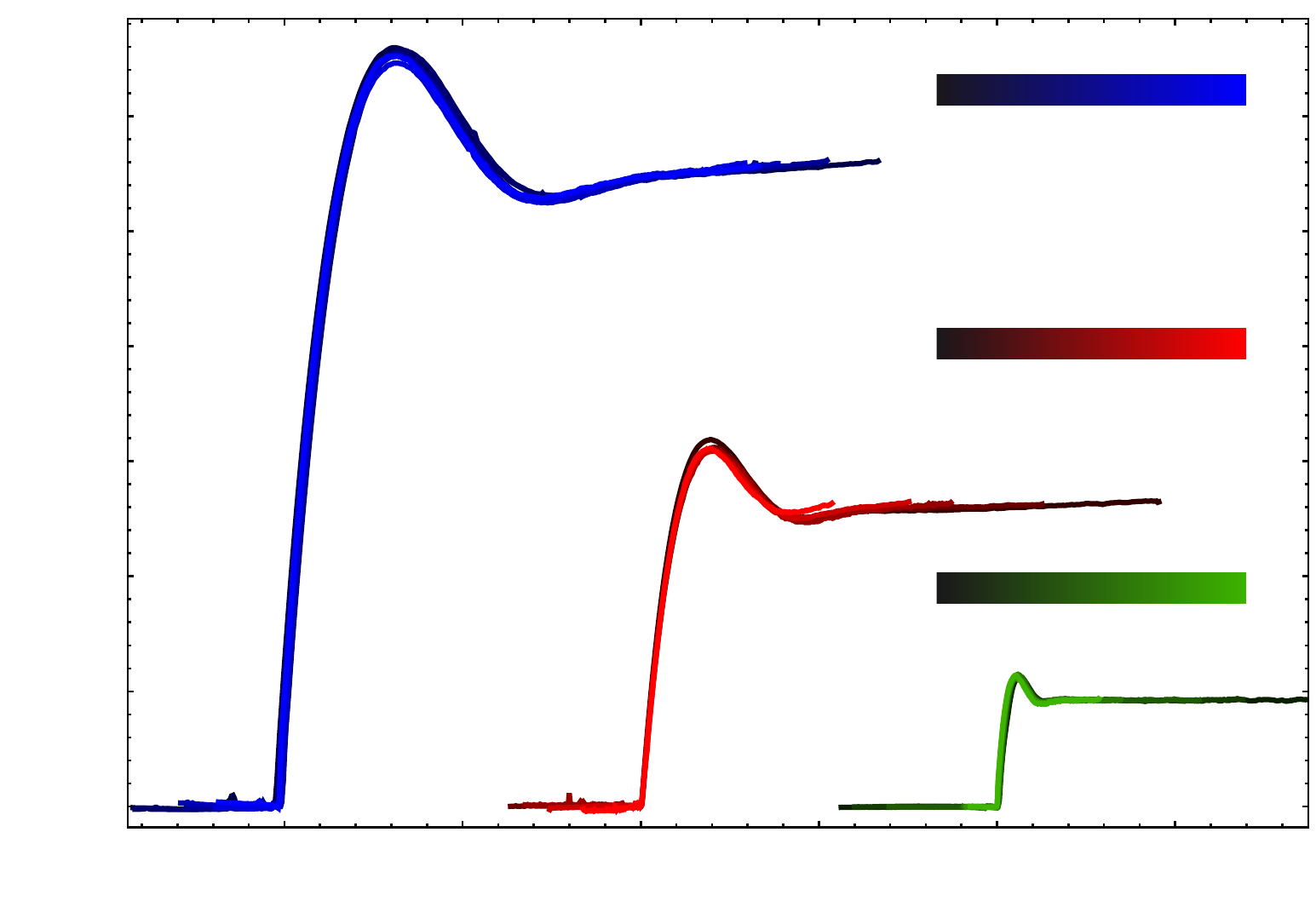}}
\put(9.6,0.7){\includegraphics[height=5.5cm]{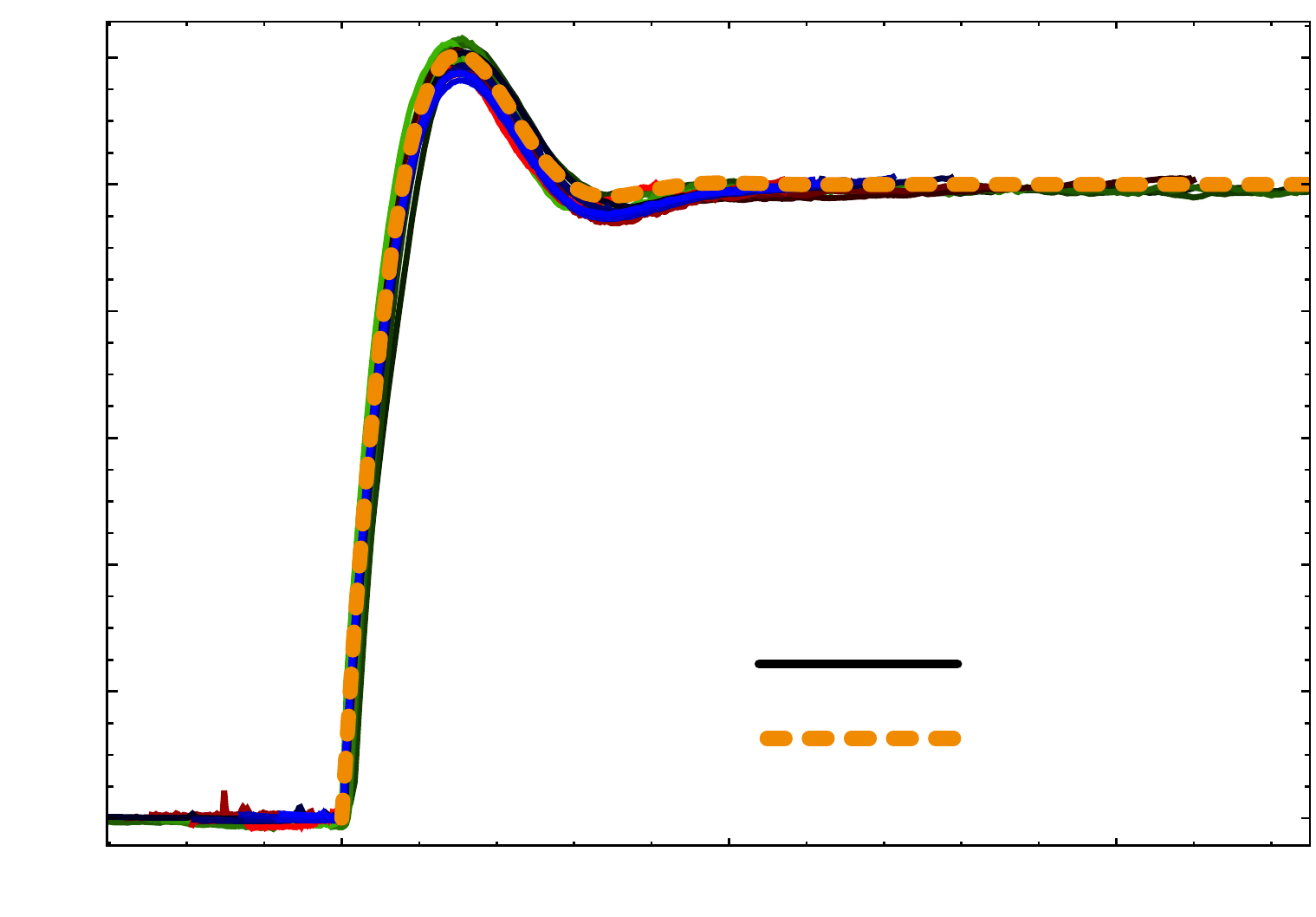}}
\put(0.1,6.11){\Large (a)}
\put(9.1,6.11){\Large (b)}
\put(0.1,2.9){\rotatebox{90}{\Large{$h \, [ \mathrm{nm}]$ }}}
\put(3,0.2){\Large{$x / t^{1/4} \, [ \, \mu \mathrm{m / min}^{1/4}\,  ]$}}
\put(2.,0.75){\fsize -10}
\put(4.4,0.75){\fsize 0}
\put(6.5,0.75){\fsize 10}
\put(0.7,5.4){\fsize 300}
\put(0.7,4){\fsize 200}
\put(0.7,2.55){\fsize 100}
\put(1.1,1.15){\fsize 0}
\put(11.9,0.15){\Large{$ \left( 3\eta / \gamma \; {h_0}^3 \right)^{1/4} \, x /  t^{1/4} $}}
\put(9.1,3.12){\rotatebox{90} {\Large{$h / h_0$ }}}
\put(9.65,5.75){\fsize 1.2}
\put(9.95,4.96){\fsize 1}
\put(9.65,4.2){\fsize 0.8}
\put(9.65,3.45){\fsize 0.6}
\put(9.65,2.7){\fsize 0.4}
\put(9.65,1.95){\fsize 0.2}
\put(9.95,1.2){\fsize 0}
\put(11.55,0.75){\fsize 0}
\put(13.8,0.75){\fsize 10}
\put(16.1,0.75){\fsize 20}
\put(15.53,2.18){\small Experiments}
\put(15.53,1.7){\small Numerics}
\put(6.9,5.27){\small $t \; \left[ \mathrm{min} \right]$}
\put(6.3,5.27){\small 0}
\put(8.,5.27){\small 90}
\put(6.9,3.7){\small $t \; \left[ \mathrm{min} \right]$}
\put(6.3,3.7){\small 0}
\put(7.9,3.7){\small 140}
\put(6.9,2.25){\small $t \; \left[ \mathrm{min} \right]$}
\put(6.3,2.25){\small 0}
\put(7.9,2.25){\small 400}
\end{picture}
\end{center}
\caption{(a) Self-similar profiles for each of the experiments shown in Fig.~\ref{fig:summary} as obtained by plotting the vertical position $h$ of the interface as a function of $x / t^{1/4}$. The experiments are shifted horizontally for clarity. 
(b) A universal profile appears when the vertical axis is rescaled by $h_0$ and the horizontal axis is non-dimensionalized by applying a lateral stretching in each experiment. The numerical solution (orange dashed line) is in excellent agreement with the experiments.}
\label{fig:universal_profile}
\end{figure*}

In Fig.~\ref{fig:universal_profile} (a), the horizontal axis is rescaled by applying the transformation $x \to x / t ^{1/4}$ and we observe that in each experiment the different profiles collapse on a single curve. This collapse demonstrates that the self-similar dynamics $h (x,t) = h (x / t^{1/4})$ is valid even for an interface with contact line, provided that the system is in the SCL regime where the line does not move. 
 
In the following the experimental profiles are compared to the numerical solution of the thin film equation, see Eq.~(\ref{eq:tfe}). The numerical profile is computed using a finite difference method \citep{bertozzi1998, salez2012numerical}. The dimensionless self-similar variable $X / T^{1/4}$ is introduced, where $X = x / \ell_0 $ and $T = \gamma \, t {h_0}^3 / (3 \eta {\ell_0}^4)$, and thus: 
\begin{equation}
\frac{X}{T^{1/4}} = \left( \frac{3 \eta}{\gamma \, {h_0}^3} \right)^{1/4 }\frac{x}{t^{1/4}} \;.
\label{eq:num_vs_exp}
\end{equation}
In addition, the condition that the contact line is fixed at $X=0$ is enforced in the algorithm.

Thus, a general picture can be obtained by rescaling the vertical axis of the experimental profiles with the initial thickness, i.e.\ $h \to h / h_0$, and by stretching the horizontal axis $ x \to (3 \eta / \gamma \, {h_0}^3)^{1/4 } \,  x / t^{1/4}$. This lateral stretch is a fitting parameter that depends on the experiment and has a clear physical interpretation \citep{mcgraw2013, salez2012numerical}. Applying this rescaling to all experimental profiles leads to a perfect collapse of all the profiles on a single master curve (Fig.~\ref{fig:universal_profile} (b)). This curve represents a universal profile of the SCL regime valid for all the parameters involved in these experiments (annealing time, film thickness, molecular weight and temperature). Note that the rescaled height of the bump is equal to $22 \pm 2 \%$ of the initial thickness of the film. 

The excellent agreement between the experiments and the thin film model (Fig.~\ref{fig:universal_profile} (b)) suggests that the thin film equation with a fixed contact line is sufficient to capture the physics of the SCL regime. The values of the resulting fitting parameters are used to compute the capillary velocity $\gamma / \eta$. Based on $\gamma = 30.8$ mN/m~\citep{brandrup1999polymer}, the viscosity $\eta$ of the PS is also evaluated (see Tab.~\ref{tab:viscosities}) and is found to be in excellent agreement with the values reported in the literature \citep{brandrup1999polymer, rubinstein2003polymer}.

\begin{table}[b!]
\begin{center}
\begin{tabular}{|  c | c | c | }
\hline 
  $M_\mathrm{w} \left[ \mathrm{kg / mol} \right]$ & $\quad T \; \left[ ^\circ \mathrm{C} \right] \quad $ & $\quad \eta \; \left[  \mathrm{Pa \; s} \right] \quad$ \\
  \hline
  3.2 & 110 & $5.8 \times 10^3$ \\
  19 & 140 & $7.8 \times 10^3$ \\
  19 & 150 & $1.6 \times 10^3$ \\
  34 & 140 & $2.9 \times 10^4$ \\
  34 & 150 & $3.1 \times 10^3$ \\ 
  \hline 
\end{tabular}
\end{center}
\caption{Viscosity $\eta$ of PS as a function of molecular weight $M_\mathrm{w}$ and annealing temperature $T$. Values of the viscosity are obtained by fitting the experimental self-similar profile of each experiment to the numerical solution, computing the capillary velocity $\gamma/\eta$ through Eq.~(\ref{eq:num_vs_exp}) and assuming $\gamma = 30.8\;$mN/m~\citep{brandrup1999polymer}.}
\label{tab:viscosities}
\end{table}

\subsection{Dewetting transition}

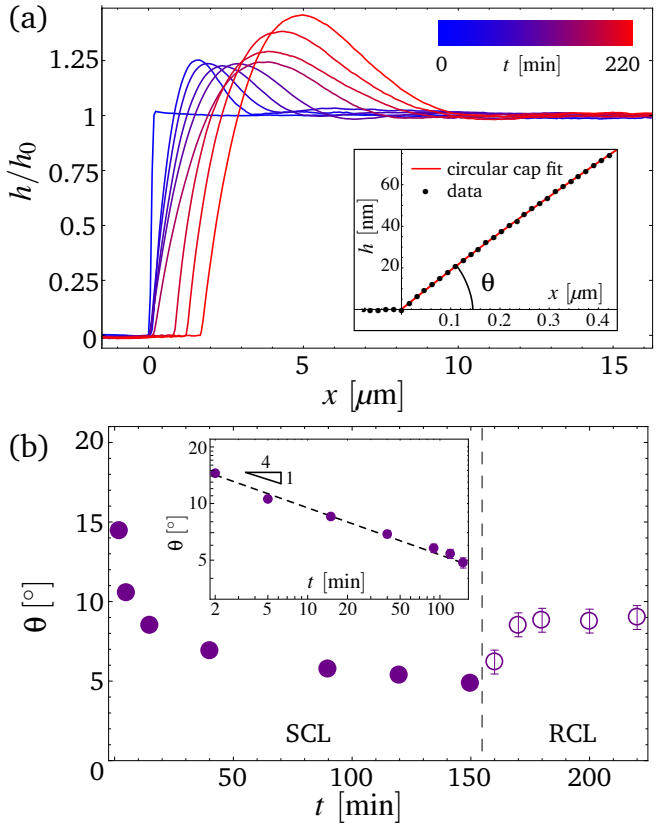
\begin{figure}[t]
\begin{center}
\setlength{\unitlength}{1cm}
\begin{picture}(8.8,11)
\put(1.2,0.7){\includegraphics[width=.868\columnwidth]{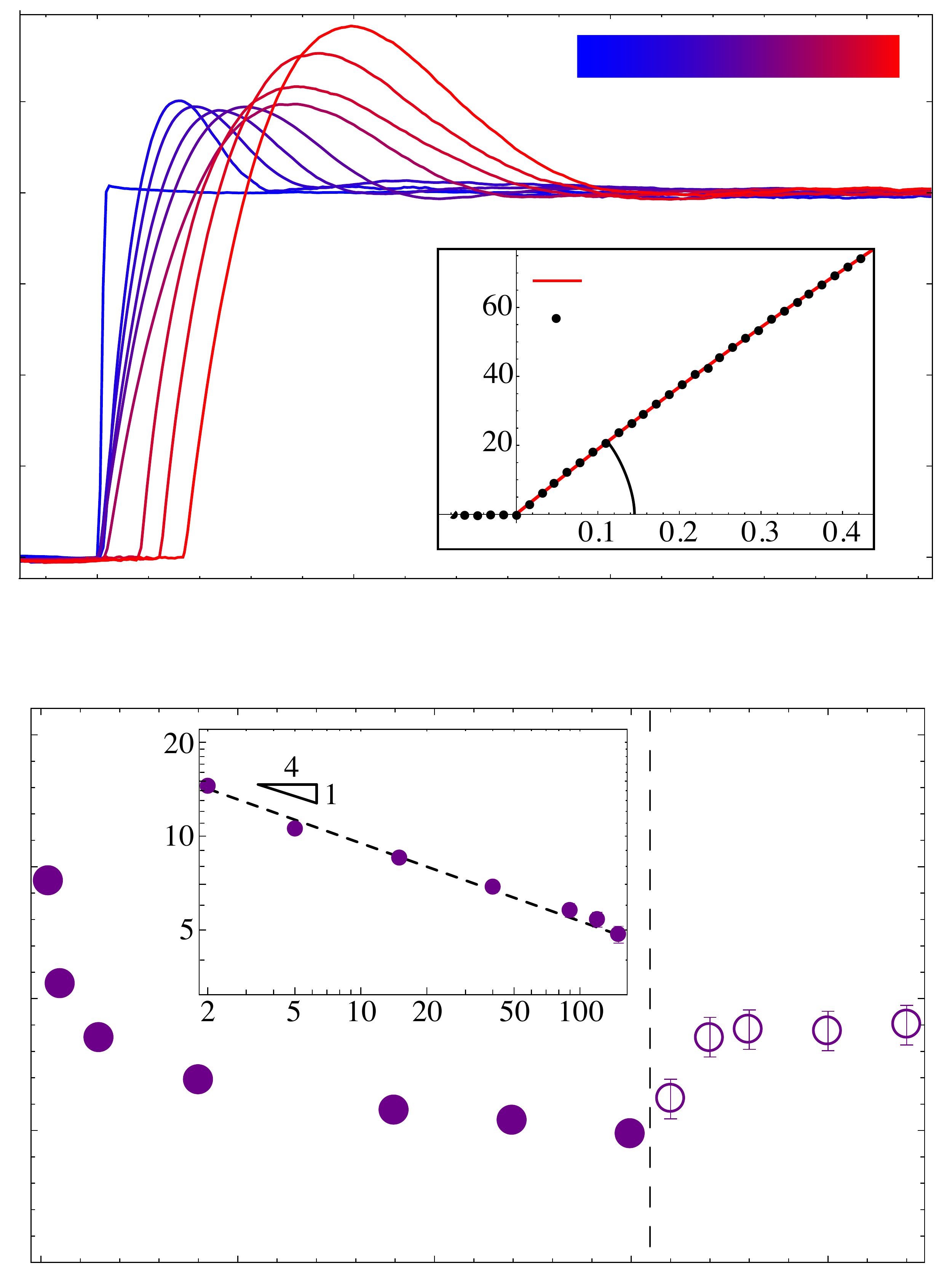}}
\put(4.2,0.13){\Large $t \; \left[ \mathrm{min} \right]$}
\put(0.3,2.55){\rotatebox{90}{\Large{$\theta \, \left[ ^\circ \right]$ }}}
\put(4.1,3.15){\small $ t \; \left[ \mathrm{min} \right]$ }
\put(2.22,3.65){\rotatebox{90}{\small{$\theta \, \left[ ^\circ \right]$ }}}
\put(1,5){\fsize 20}
\put(1,3.9){\fsize 15}
\put(1,2.85){\fsize 10}
\put(1.2,1.8){\fsize 5}
\put(1.2,0.6){\fsize 0}
\put(2.9,0.52){\fsize 50}
\put(4.38,0.52){\fsize 100}
\put(5.95,0.52){\fsize 150}
\put(7.55,0.52){\fsize 200}
\put(4.3,5.6){\Large $x \; \left[ \mathrm{\mu m} \right]$}
\put(0.1,8.3){\rotatebox{90}{\Large{$h/h_0$ }}}
\put(1.9,6){\fsize 0}
\put(3.93,6){\fsize 5}
\put(5.9,6){\fsize 10}
\put(7.98,6){\fsize 15}
\put(0.65,10.07){\fsize 1.25}
\put(1.1,9.3){\fsize 1}
\put(0.65,8.55){\fsize 0.75}
\put(0.8,7.85){\fsize 0.5}
\put(0.65,7.1){\fsize 0.25}
\put(1.1,6.35){\fsize 0}
\put(3.8,1.1){\fsize SCL}
\put(7.3,1.1){\fsize RCL}
\put(7.3,7.){\small $ x \; \left[ \mathrm{\mu m} \right]$ }
\put(4.75,7.6){\rotatebox{90}{\small $ h \; \left[ \mathrm{n m} \right]$}}
\put(6.4,7.1){\fsize $\theta$}
\put(5.95,8.65){\footnotesize circular cap fit}
\put(5.95,8.35){\footnotesize data}
\put(6.7,10.03){\small $t \; \left[ \mathrm{min} \right]$}
\put(5.8,10.03){\small 0}
\put(8.05,10.03){\small 220}
\put(0.1,10.6){\Large (a)}
\put(0.1,4.9){\Large (b)}
\end{picture}
\end{center}
\caption{
(a) Profiles corresponding to the stationary contact line (before 150\,min) and to the receding contact line (after 150\,min) regimes. As soon as the three-phase contact line recedes, the height of the bump grows. Here, $h_0 = 120$\,nm and $M_\mathrm{w} = 34$\,kg/mol at $T = 140\,^\circ$C. The inset displays the profile recorded close to the contact line ($t= 5$\,min, black points) and a circular arc fit (red line) used to calculate the contact angle. 
(b) Temporal evolution of the contact angle for the experiment shown in (a). Filled circles correspond to the SCL regime and open circles to the RCL regime. The error bar of the contact angle in the SCL regime is typically $\pm\, 0.3^\circ$ and, thus, smaller than the symbol size.}
\label{fig:dewetting}
\end{figure}

Self-similarity breaks down as soon as the dewetting transition takes place. Indeed, during the receding motion of the contact line the liquid accumulates in the region of the bump, causing its growth in width and height (Fig.~\ref{fig:dewetting} (a)). The occurrence of dewetting is a common observation in all the experiments. 
To shed light on the time scale of the onset of dewetting, the entire evolution of the contact angle $\theta$ between liquid and solid is monitored. For each profile close-up AFM scans (typically $2 - 4 \, \mu$m) around the contact line are performed. The value of the contact angle $\theta$ is then calculated from these profiles by fitting the shape of the liquid interface with a circular arc, as illustrated in the inset of Fig.~\ref{fig:dewetting} (a), and taking the tangent to the circle at the contact line. The equilibrium contact angle $\theta_Y$ for PS on a Si wafer has also been evaluated by annealing the film for very long time until isolated droplets form. The value $\theta_Y = 10^\circ \pm 2^\circ$ has been obtained, which is in close agreement to the one reported earlier in \citet{seemann2001polystyrene} for the same system. 

Fig.~\ref{fig:dewetting} (b) shows a typical evolution of the the contact angle. 
During the SCL regime the angle $\theta$ monotonically decreases due to the relaxation of the interface and to the stationary position of the three-phase contact line line. Interestingly, the SCL regime extends even when the angle is smaller than $\theta_\mathrm{Y}$. 
Eventually, the receding motion of the line takes place when the angle reaches a critical value $\theta^* < \theta_\mathrm{Y}$. As soon as the contact line retracts, $\theta$ rapidly increases to a roughly constant receding contact angle of the moving front. 

In all the experiments $\theta$ decreases following a $t^{-1/4}$ power law (see Fig.~\ref{fig:dewetting} (b), inset) in the SCL regime.  This power law is a direct consequence of the self-similar evolution of the liquid interface. Indeed, as shown above, the horizontal length scales evolve as $\sim t^{1/4}$, which implies for a constant vertical length scale $h_0$ that $\tan\theta\sim t^{-1/4}$, and thus $\theta\sim t^{-1/4}$ at small angles. The $\theta \sim t^{-1/4}$ relation leads to a fast decrease of the angle at short time, so that it generally drops to small values in a few minutes.

Let us now  introduce the dimensionless time $\tau = t\gamma / (h_0\eta)$. Using the values of $h_0$ recorded with the AFM and those of  $\eta/\gamma $ extracted from the comparison between the self-similar profile and the numerical solution, $\theta$ is plotted as a function of $\tau$ (see Fig.~\ref{fig:universal_angles}). The values of the contact angles for all the experiments collapse on the same master curve, a significant result because the experiments involve different heights of the film and  different capillary velocities, since the viscosity changes by more than one order of magnitude. An important observation is the fact that in all the experiments the retraction of the line precisely appears at the same value of the contact angle $\theta^* = 4.5^\circ \pm 0.5^\circ<\theta_\mathrm{Y}$. From the master curve we, hence, deduce that the dimensionless dewetting time is also universal and equal to $\tau^* \simeq 10^5$. 
After the dewetting transition an exponential relaxation of the contact angle might be expected, although the experimental uncertainty of the contact angle measurement together with the time resolution here can not provide a precise validation. 
Note that the rescaling with $\tau$ satisfied by the SCL regime does not necessarily apply for the RCL regime. 

\section{Discussion}


From these results it appears that the stationary position of the contact line followed by the onset of dewetting are general and robust features and can not be attributed to the presence of randomly distributed pinning sites, i.e.\ local topographical defects and / or chemical inhomogeneities: for all seven experiments  and all annealing times the invariance of the profiles in the third dimension as well as a perfect symmetric representation of the profiles throughout the relaxation process (see Fig.~\ref{fig:stripe}) are valid. The universal transition between both regimes is reproducible and appears to be independent of the liquid parameters after proper rescaling of the relaxation dynamics. In the following paragraph, we briefly discuss the influence of the substrate on the dynamics.  

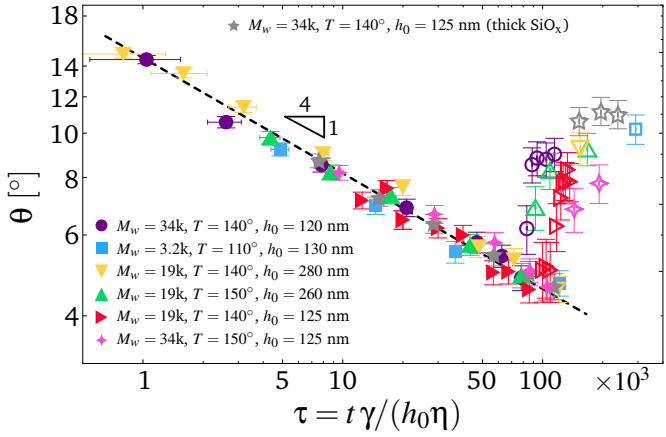
\begin{figure}[t]
\begin{center}
\setlength{\unitlength}{1cm}
\begin{picture}(8.8,5.8)
\put(0.75,0.55){\includegraphics[width=.92\columnwidth]{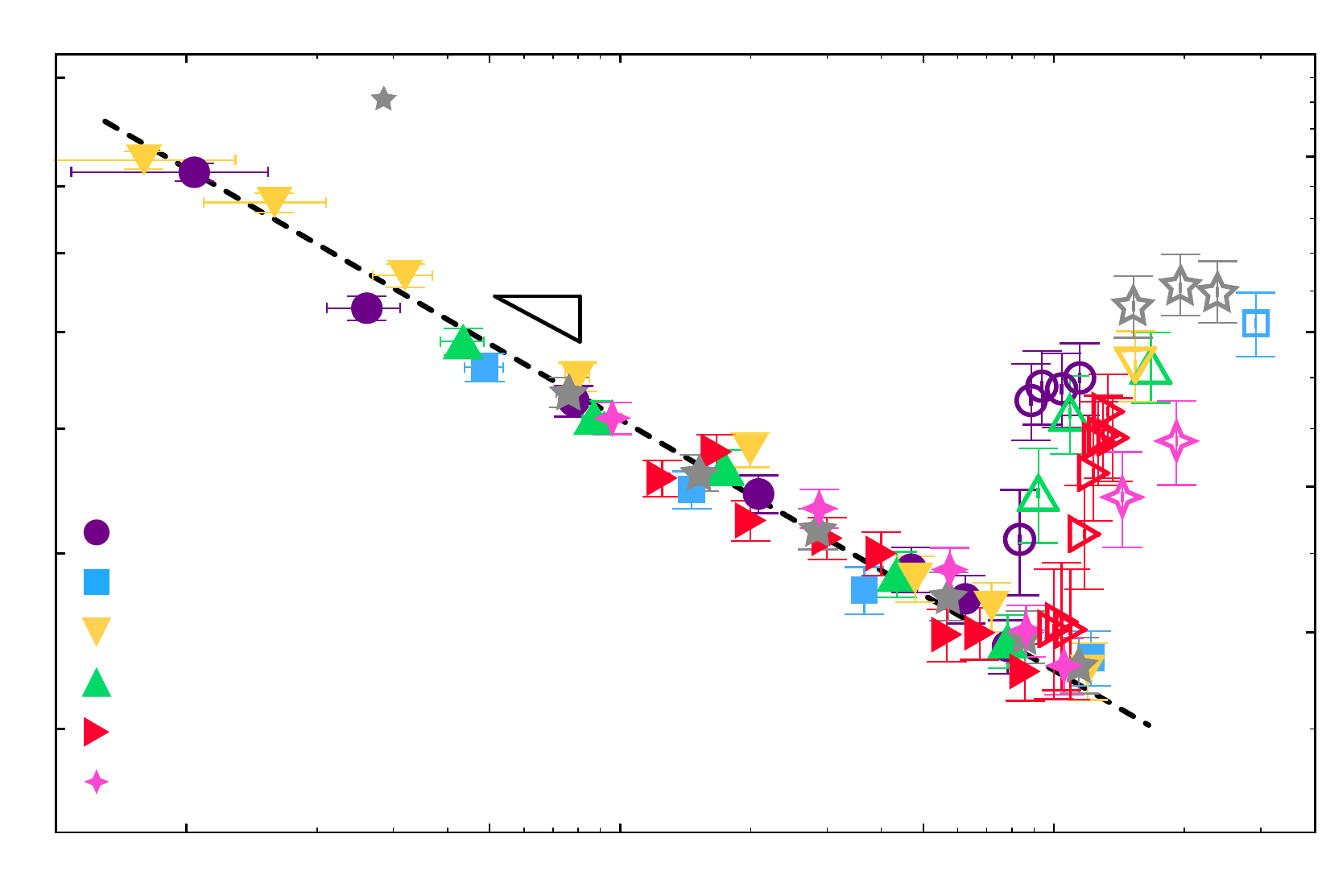}}
\put(3.9,0.15){\Large{$\tau = t \, \gamma / (h_0 \eta)$}}
\put(0.1,2.8){\rotatebox{90}{\Large{$\theta \, \left[ ^\circ \right]$ }}}
\put(1.55,2.7){\scriptsize $M_w = 34$k, $T = 140^\circ$, $h_0 = 120$ nm}
\put(1.55,2.4){\scriptsize $M_w = 3.2$k, $T = 110^\circ$, $h_0 = 130$ nm}
\put(1.55,2.1){\scriptsize $M_w = 19$k, $T = 140^\circ$, $h_0 = 280$ nm}
\put(1.55,1.8){\scriptsize $M_w = 19$k, $T = 150^\circ$, $h_0 = 260$ nm}
\put(1.55,1.5){\scriptsize $M_w = 19$k, $T = 140^\circ$, $h_0 = 125$ nm}
\put(1.55,1.2){\scriptsize $M_w = 34$k, $T = 150^\circ$, $h_0 = 125$ nm}
\put(3.3,5.37){\scriptsize $M_w = 34$k, $T = 140^\circ$, $h_0 = 125$ nm (thick SiO$_\mathrm{x}$)}
%
%
\put(0.65,5.45){\fsize 18}
\put(0.65,4.77){\fsize 14}
\put(0.65,4.35){\fsize 12}
\put(0.65,3.87){\fsize 10}
\put(0.85,3.28){\fsize 8}
\put(0.85,2.53){\fsize 6}
\put(0.85,1.45){\fsize 4}
\put(1.82,0.6){\fsize 1}
\put(3.65,0.6){\fsize 5}
\put(4.35,0.6){\fsize 10}
\put(6.2,0.6){\fsize 50}
\put(6.88,0.6){\fsize 100}
\put(7.88,0.59){\fsize $\times 10^3$}
\put(3.97,4.3){\small 4}
\put(4.35,3.98){\small 1}
\end{picture}
\end{center}
\caption{Contact angle $\theta$ plotted as a function of dimensionless time $\tau$ for all experiments. Data for the SCL regime are plotted with filled symbols while data for RCL regime are plotted with open symbols. 
All the data for the SCL regime follow the $\tau ^{-1/4}$ power law and collapse on the same master curve. 
The transition to dewetting takes place at the same values of $\theta$ and $\tau$ for all the experiments.}
\label{fig:universal_angles}
\end{figure}

Aside from the general features outlined above, a prominent and interesting observation is the fact that a stationary position of the contact line is observed for values of the contact angle smaller than the equilibrium one, but larger than the critical one, i.e.\ $\theta^* <\theta< \theta_\mathrm{Y}$. 
This observation has been confirmed even for scan sizes in proximity of the contact line and in the presence of small defects that can be used as reference points, although the limited lateral resolution of the AFM can not accurately detect displacements in the order of a few nanometers. 
The dynamics of thin liquid films is governed by short-range as well as long-range forces, e.g.\ originating from van-der-Waals interactions, between the substrate and the liquid. In principle these long-range forces are negligible for a film thickness larger than $\sim 30$\,nm \citep{seemann2001dewetting}, which is always the case in the experiments discussed here. However, in the region close to the three-phase contact line the thickness of the film decreases monotonically to zero and the extent of the zone where $h < 30$\,nm grows as soon as the contact angle decreases. In previous work it has already been shown that long-range intermolecular forces might affect the shape of the liquid interface close to the contact line of nanometric droplets \citep{seemann2001polystyrene}. In order to test if long-range forces play a role in this zone and in particular if they affect the onset of the retraction of the line experiments on Si wafers exhibiting a thick ($150$\,nm) oxide layer have been carried out. This choice is motivated by the fact that the presence of a thick oxide layer considerably changes the effective interface potential comprising short- and long-range forces and represents a well-established model system\citep{seemann2001dewetting, seemann2001polystyrene}. One separate experiment has been performed on thick oxide layer Si wafers (see Fig.~\ref{fig:universal_angles}), and it has not shown any significant difference with respect to the experiments on a native Si oxide layer: the values of the contact angles for this experiment perfectly collapse on the master curve in Fig.~\ref{fig:universal_angles} and the contact angle at the transition is preserved. Hence, we conclude that long-range forces affect neither the relaxation dynamics nor the onset of motion of the contact line. A possible alternative explanation would be the presence of a contact angle hysteresis related to a uniform intrinsic pinning potential of the polymer molecules on the Si wafers, yet to be substantiated in future experimental work. 



The occurrence of dewetting, despite the fact the initial contact angle is much larger than the equilibrium one, is a common feature in all the experiments and might appear counterintuitive at first glance. Indeed in a liquid interface with constant curvature (a spherical drop, or a cylindrical drop in the 2D case) the advancing or receding motion of the contact line can be ultimately predicted from the value of the contact angle: In particular the situation $\theta_0 > \theta_Y$, which characterizes all our experiments, would have lead to the monotonic spreading of the liquid. In fact it is easy to prove that a spherical or cylindrical interface in the absence of gravity reaches its minimum of energy when the forces at the contact line are at equilibrium, which is precisely the foundation of Young's construction of the equilibrium angle. However, in the more general case of a non-constant curvature interface, the equilibrium of the forces at the contact line given by $\theta = \theta_Y$ does not correspond to the minimum of the energy anymore. The system has to adjust the contact angle \textit{and} to relax the liquid interface at the same time in order to achieve the global minimum and, thus, it is not possible to predict spreading or dewetting \textit{ab initio} only from the value of the contact angle.

A simple geometrical argument based on the comparison of the initial state and the final state of the system can be introduced to anticipate the occurrence of spreading or dewetting. In the 2D configuration, the initial state is a rectangular interface having width $2 \ell_0$ and thickness $h_0$ (see Fig.~\ref{fig:stripe} (b)). The final state is a circular cap having contact line radius $r$ and equilibrium contact angle $\theta_Y$. Invoking volume (area, in 2D) conservation between the two states, the contact line radius can be deduced as a function of $h_0$, $\ell_0$ and $\theta_Y$. We define the wetting parameter $\mathcal{W} = {r}/{\ell_0}$ and show that:
\begin{equation}
\mathcal{W} =  \sqrt{\frac{2h_0}{\ell_0}} \frac{\sin \theta_Y}{\sqrt{\theta_Y - \sin \theta_Y \cos \theta_Y}} \;.
\end{equation} 
A spreading or dewetting situation is triggered by $\mathcal{W} > 1$ and  $\mathcal{W} < 1$, respectively. Note that $\mathcal{W}$ only depends on the aspect ratio of the initial stripe $\ell_0 / h_0$ and the equilibrium contact angle. For the experiment illustrated in Fig.~\ref{fig:stripe} (a), the aspect ratio is $\ell_0/h_0 \simeq 167$ and the wetting parameter that corresponds to the experimental value $\theta_Y = 10^\circ$ is $\mathcal{W} = 0.32$ . Hence, dewetting appears to be an inevitable consequence of the initial geometry.

\section{Conclusions}

In this article we have studied the relaxation dynamics of the contact angle between a viscous liquid and a smooth solid substrate. The temporal evolution of a liquid nanofilm in the presence of three-phase contact lines has been monitored for different film thicknesses and liquid viscosities. In all experiments the initial regime, defined by a stationary position of the contact line, is followed by a second one in which dewetting takes place. We have shown that the stationary contact line regime can be described in terms of levelling dynamics in which the liquid profile exhibits a self-similar evolution in excellent agreement with a numerical solution of the thin film equation. In this regime the energy of the system diminishes due to the relaxation of the curvature of the interface and the contact angle follows a characteristic power law. The self-similarity breaks down as soon as the contact line retracts.  The transition between stationary and receding regimes of the contact line is triggered for a critical angle $\theta^* < \theta_\mathrm{Y}$ that is independent of the molecular weight of the polymer, viscosity, film thickness as well as long-range interactions. A universal transition has then emerged in terms of a characteristic dimensionless time. 

In future work, we envision to explore whether and how the robust features observed in our experiments can be generalized to different types of substrates exhibiting different surface energies and/or a variation of the hydrodynamic boundary condition between liquid and solid. These experiments might provide new fundamental insights on the contact-line dynamics at the nanoscale.

\section{Acknowledgement}
The authors would like to thank Kari Dalnoki-Veress, Joshua D. McGraw and Stephan Herminghaus for insightful discussions. The German Research Foundation (DFG) is acknowledged for financial support under grant BA 3406/2.\\





\bibliography{biblio} 
\bibliographystyle{rsc} 

\end{document}